\documentclass[manuscript,screen]{acmart}

\setcopyright{acmcopyright}
\copyrightyear{2025}
\acmYear{2025}
\usepackage[per-mode=symbol,per-symbol=p]{siunitx}
\usepackage[utf8]{inputenc}
\usepackage{adjustbox}
\usepackage{hyperref} 
\usepackage{multirow}
\usepackage[ruled,vlined]{algorithm2e}
\usepackage{algorithmic}
\usepackage{graphicx}
\usepackage{textcomp}
\usepackage{xcolor}
\usepackage{balance}
\usepackage{enumitem}
\usepackage{setspace}
\usepackage{wrapfig}
\usepackage{listings}
\usepackage{color}
\usepackage{caption}
\usepackage{subcaption}
\usepackage{makecell}

\usepackage{ctable}
\usepackage{balance}

\newcommand{\cf}{\emph{cf.}\xspace}
\newcommand{\etal}{\emph{et al.}\xspace}
\newcommand{\ie}{\emph{i.e.}, }
\newcommand{\eg}{\emph{e.g.}, }

\newcommand{\HAS}{\emph{HTTP Adaptive Streaming }}

\newcommand{\HEVC}{\emph{High Efficiency Video Coding }}

\newcommand{\VVC}{\emph{Versatile Video Coding }}

\newcommand{\UHD}{\emph{Ultra High Definition }}

\newcommand{\QoE}{\emph{Quality of Experience}\xspace}

\newcommand{\DynResXPSNR}{\texttt{DynResXPSNR}\xspace}
\newcommand{\DynResVMAF}{\texttt{DynResVMAF}\xspace}
\newcommand{\vexus}{\texttt{VEXUS}\xspace}
\newcommand{\jtps}{\texttt{JTPS}\xspace}
\newcommand{\jqtpf}{\texttt{JQT-PF}\xspace}
\newcommand{\jrqtpf}{\texttt{JRQT-PF}\xspace}
\newcommand{\default}{\texttt{Default}\xspace}
\newcommand{\fixedladder}{\texttt{Fixedladder}\xspace}

\newcommand{\vig}[1]{\textcolor{black}{#1}}
\newcommand{\rev}[1]{\textcolor{black}{#1}}

\newcommand{\EY}{$E_{\text{Y}}$}

\newcommand{\LY}{$L_{\text{Y}}$}

\newcommand{\h}{$h$}

\newcommand{\BDRP}{BDR\textsubscript{P}\xspace}
\newcommand{\BDRV}{BDR\textsubscript{V}\xspace}
\newcommand{\BDRX}{BDR\textsubscript{X}\xspace}

\begin{document}

% \title{Exploring Quality of Experience Factors for Efficient and Sustainable Adaptive VVC Streaming}
\title{Multi-Objective Pareto-Front Optimization for Efficient Adaptive VVC Streaming}

\author{Angeliki Katsenou}
%\email{Amritha.Premkumar@b-tu.de}
\orcid{}
%\authornotemark[1]
\affiliation{%
  \institution{Visual Information Lab}
  \institution{School of Computer Science}
  \institution{University of Bristol} 
  \streetaddress{}
  %\city{Cottbus}
  %\state{Brandenburg}
  \country{UK}
  %\postcode{03046}
}

\author{Vignesh V Menon}
\email{vignesh.menon@hhi.fraunhofer.de}
\orcid{0000-0003-1454-6146}
\affiliation{
  \institution{Video Communication and Applications Dept}
  \institution{Fraunhofer HHI}
  \city{Berlin}
  \country{Germany}
}

\author{Guoda Laurinaviciute}
%\email{Amritha.Premkumar@b-tu.de}
\orcid{}
%\authornotemark[1]
\affiliation{%
  \institution{Visual Information Lab}
  \institution{School of Computer Science}
  \institution{University of Bristol} 
  \streetaddress{}
  %\city{Cottbus}
  %\state{Brandenburg}
  \country{UK}
  %\postcode{03046}
}

\author{Benjamin Bross}
\email{vignesh.menon@hhi.fraunhofer.de}
\orcid{0000-0003-1454-6146}
\affiliation{
  \institution{Video Communication and Applications Dept}
  \institution{Fraunhofer HHI}
  \city{Berlin}
  \country{Germany}
}

\author{Detlev Marpe}
\email{vignesh.menon@hhi.fraunhofer.de}
\orcid{0000-0003-1454-6146}
\affiliation{
  \institution{Video Communication and Applications Dept}
  \institution{Fraunhofer HHI}
  \city{Berlin}
  \country{Germany}
}

\renewcommand{\shortauthors}{A. Katsenou et al.}

\begin{abstract}
\vig{Adaptive video streaming has facilitated improved video streaming over the past years. A balance among coding performance objectives such as bitrate, video quality, and decoding complexity is required to achieve efficient, content- and codec-dependent, adaptive video streaming. This paper proposes a multi-objective Pareto-front (PF) optimization framework to construct quality-monotonic, content-adaptive bitrate ladders \VVC (VVC) streaming that jointly optimize video quality, bitrate, and decoding time, which is used as a practical proxy for decoding energy. Two strategies are introduced: the Joint Rate-Quality-Time Pareto Front (\jrqtpf) and the Joint Quality-Time Pareto Front (\jqtpf), each exploring different tradeoff formulations and objective prioritizations. The ladders are constructed under quality monotonicity constraints during adaptive streaming to ensure a consistent Quality of Experience (QoE). Experiments are conducted on a large-scale UHD dataset (Inter-4K), with quality assessed using PSNR, VMAF, and XPSNR, and complexity measured via decoding time and energy consumption. The \jqtpf method with tuning parameter $\alpha_{J} = 2.5$ achieves \SI{11.76}{\percent} average bitrate savings while reducing average decoding time by \SI{0.29}{\percent} to maintain the same XPSNR, compared to a widely-used fixed ladder. More aggressive configurations ($\alpha_{J} = 0.5$) yield up to \SI{27.88}{\percent} bitrate savings at the cost of increased complexity. The \jrqtpf strategy, on the other hand, offers more controlled tradeoffs, achieving \SI{6.38}{\percent} bitrate savings and \SI{6.17}{\percent} decoding time reduction for tuning parameter $\alpha_{M} = 0.75$. This framework outperforms existing methods, including fixed ladders, VMAF- and XPSNR-based dynamic resolution selection, and complexity-aware benchmarks. The results confirm that PF optimization with decoding time constraints enables sustainable, high-quality streaming tailored to network and device capabilities.}
% Content-tailored methods improve the quality of experience at the cost of higher decoding complexity. By leveraging multi-objective Pareto-front optimization, this paper investigates the construction of optimized bitrate ladders for . Experimental results over a large corpus of video sequences validate the advantages of the proposed parameterisable multi-objective optimization framework, showcasing improved quality consistency, bitrate savings, decoding efficiency, and energy savings over state-of-the-art approaches focusing primarily on quality improvement.
\end{abstract}




\maketitle

\section{Introduction}
As the vast volume of data traffic over the internet concerns video streaming services~\cite{sandvine2023}, research efforts have been focused on more effective compression technologies. However, at each generation of video coding standards, the compression performance is improved over the previous generations at the cost of computational complexity~\cite{Ronca}. The transition from \HEVC~(HEVC)~\cite{HEVC} to \VVC~(VVC)\cite{vvc_ref}, the latest video coding standard, has been driven by the substantial improvements in compression efficiency offered by VVC~\cite{hevc_vvc_enc_comp}. These advances are crucial to delivering high-resolution content efficiently. Despite these benefits, the increased computational complexity of VVC presents challenges, particularly for energy-constrained client devices\cite{vvc_complexity,hevc_vvc_enc_comp,dec_energy_vvc}. Although decoding complexity is relatively small compared to encoding, it still poses a challenging computational and energy-draining task on battery-constrained user devices. 

Effective compression alone is insufficient for achieving high \QoE~(QoE) in video content distribution. Video service providers must enable the delivery of high-quality content at scale, which entails heterogeneous networks (types and speeds) and display devices. For delivering the on-demand encoded video content, \HAS~(HAS) is the dominating standard. It dynamically adapts the quality of the video stream according to network conditions and user device specifications, delivering an optimal viewing experience under diverse scenarios and for all types of content~\cite{HAS_survey}. This dynamic adaptation is enabled by encoding video content at various bitrates and spatiotemporal resolutions (different video representations), forming a bitrate ladder from which the most suitable representation is selected in real-time for each user~\cite{mpeg_dash_ref,farahani2024ai_sustainable}. As a result, designing an efficient bitrate ladder is critical for HAS, as it significantly affects the user's QoE and decoding complexity.

Traditional ladder construction methods often rely on fixed~\cite{HLS_ladder_ref} or heuristic-based approaches, primarily focused on rate-quality (RQ) curves that do not adequately balance the trade-offs between bitrate, video quality, and decoding complexity~\cite{netflix_paper, Katsenou_IEEEOJSP2021, jnd_streaming, ensemble_learning_vvc_ladder, Menon_ICME2022, menon2024convexhull_xpsnr, Telili_ref}. Pareto-front (PF) optimization offers a robust framework for addressing these trade-offs. By identifying a set of optimal solutions representing the best possible compromises between conflicting objectives, PF optimization techniques enable the construction of bitrate ladders that maximize video quality while minimizing bitrate and decoding complexity. Several quality metrics can be employed in this context, such as Peak Signal-to-Noise Ratio(PSNR)~\cite{Katsenou_IEEEOJSP2021}, eXtended PSNR (XPSNR)~\cite{XPSNR, wien_xpsnr_vs_vmaf} and Video Multi-Method Assessment Fusion (VMAF)~\cite{VMAF,Katsenou_vmaf_ladder}. %However, there has been evidence that XPSNR aligns better perceptually compared to other quality metrics for VVC encodes, thus it can predict the subjective codec ranking reported in~\cite{wien_xpsnr_vs_vmaf} with improved accuracy compared to VMAF~\cite{itu_xpsnr_vs_vmaf}.

\begin{figure*}[t]
\centering
\begin{subfigure}{0.325\linewidth}
    \centering
    \includegraphics[width=\textwidth]{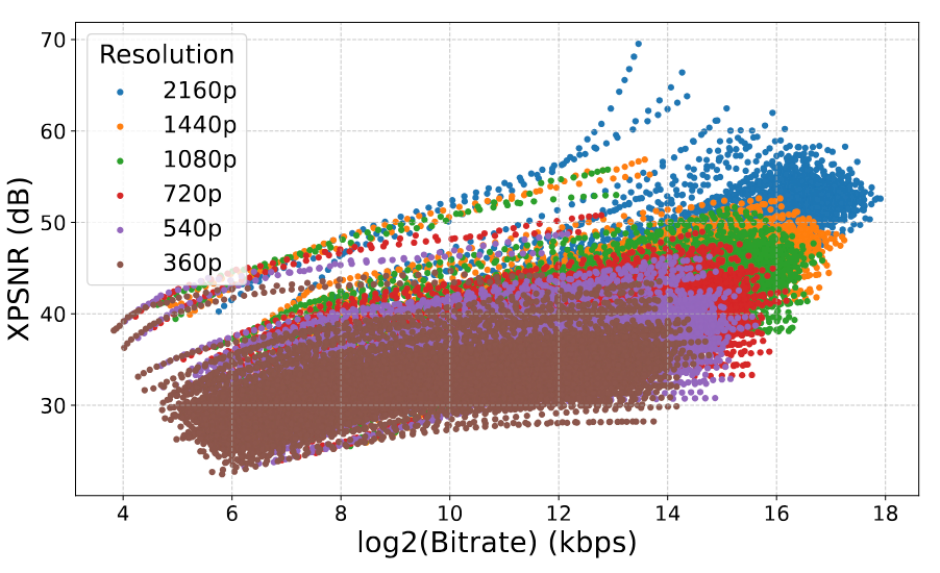}
    \caption{ Rate-Quality.}
    \label{fig:rate_quality}
\end{subfigure}
\begin{subfigure}{0.325\linewidth}
    \centering
    \includegraphics[width=\textwidth]{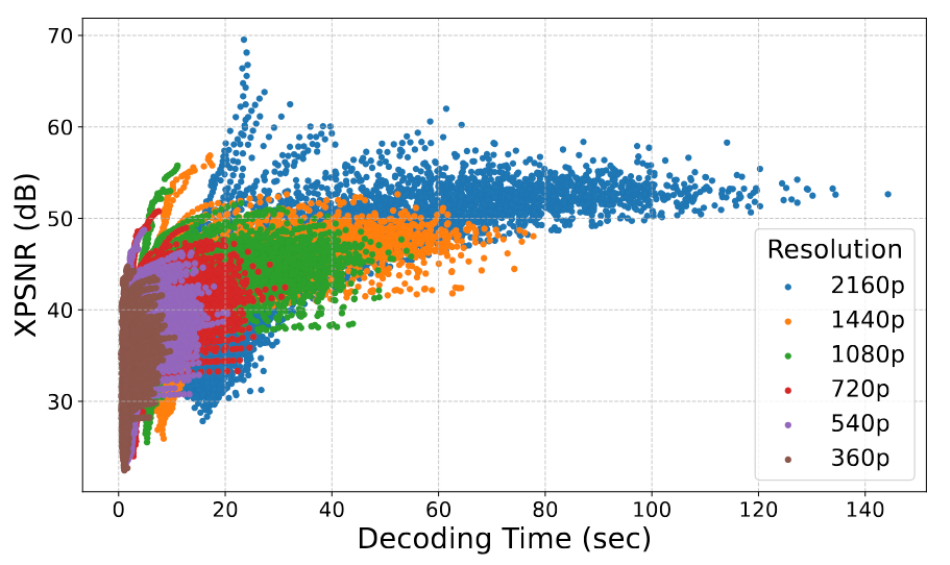}
    \caption{ Decoding time-Quality.}
\end{subfigure}
\begin{subfigure}{0.325\linewidth}
    \centering
    \includegraphics[width=\textwidth]{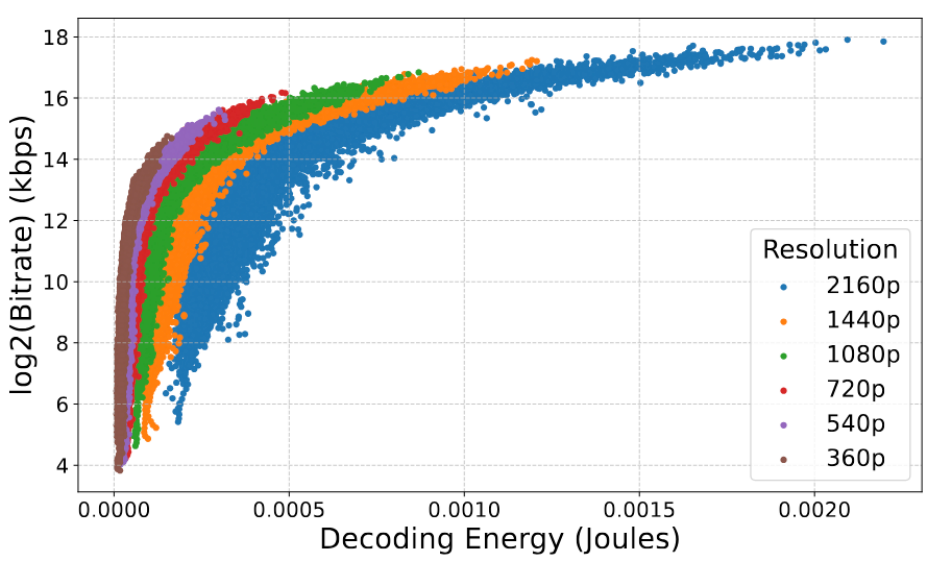}
    \caption{ Rate-Decoding energy.}
\end{subfigure}

\caption{Quality–Rate–Time–Energy points for VVenC/VVdeC~\cite{vvenc_ref, VVdeC_ref} across spatial resolutions for a quartile of the Inter-4K dataset~\cite{inter4k_ref}. Decoding time and energy consumption were measured \rev{using the Running Average Power Limit (RAPL) interface and the \texttt{CodeCarbon} tool~\cite{codecarbon_ref} on a Dell laptop (Intel Core i7-12800H processor, 32 GB RAM).}}
\label{fig:ParamSpace}
\end{figure*}

Although mixed-codec streaming environments (\eg HEVC~\cite{HEVC}, VP9~\cite{VP9}, AV1~\cite{AV1}) are typical in practice, this study focuses on VVC. VVC is the most recent video coding standard, particularly suited for \UHD and high-frame-rate content, where computational trade-offs are more critical. 
\rev{Furthermore, based on recent evaluations from ISO/IEC SC29/AG5~\cite{cvqm_phase1,cvqm_phase1b} report that both VMAF and XPSNR achieve comparable correlation with subjective quality for VVC-coded sequences. While VMAF is an industry standard, it is computationally expensive. In contrast, XPSNR offers comparable performance at a fraction of the cost, making it suitable for large-scale studies of ladder construction. Therefore, in this work, we evaluate our optimization framework under both VMAF and XPSNR.}

\begin{figure}[t]
\centering
\begin{subfigure}{0.425\linewidth}
    \centering
    \includegraphics[width=\textwidth]{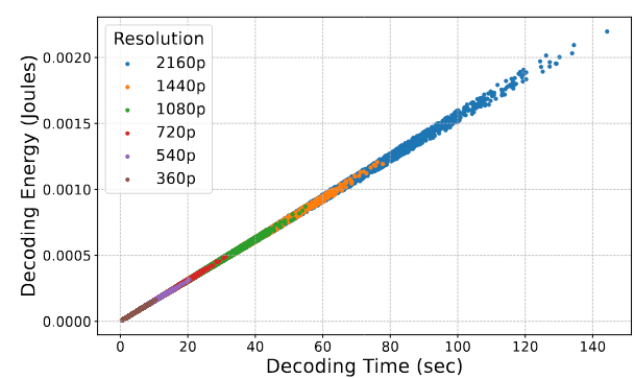}
    \caption{Dell laptop.}
    \label{fig:time_energy1}
\end{subfigure}
\begin{subfigure}{0.425\linewidth}
    \centering
    \includegraphics[width=\textwidth]{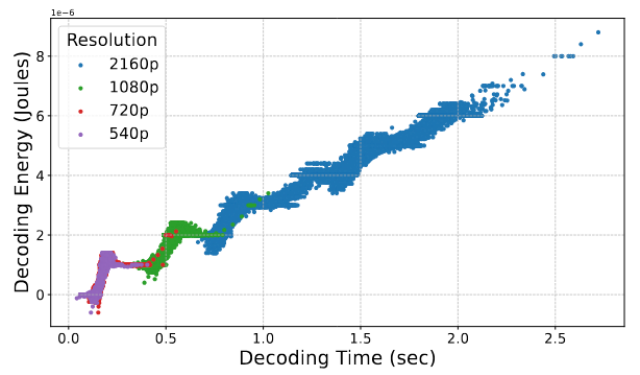}
    \caption{Mac Mini.}
    \label{fig:time_energy2}
\end{subfigure}
\caption{Decoding Time–Energy points for VVdeC~\cite{VVdeC_ref} across spatial resolutions for a quartile of the Inter-4K dataset for Dell laptop and Mac Mini.}
\label{fig:energy_time_corr}
\end{figure}

This paper explores using decoding time as a proxy to optimize the decoding \vig{complexity}, as this approach offers several advantages over \vig{metrics such as} decoding energy \vig{consumption} measurements. One of the primary benefits is its direct relevance to user experience. Decoding time is a more immediate and perceptible factor for end users, directly affecting playback smoothness and latency~\cite{dec_time_energy_rel}. Streaming services can achieve faster startup times and reduced buffering by optimizing decoding, thereby significantly enhancing QoE. Additionally, decoding time is a more straightforward metric to measure and optimize than decoding energy, which can vary widely across devices, usage patterns, and battery health. This variability makes energy measurements less consistent and more challenging to standardize across different devices. 
Moreover, focusing on decoding time saves energy, as shorter processing times typically require less power. This pseudo-linear relationship between decoding time and energy is observed in Fig.~\ref{fig:time_energy1}, where the Pearson correlation coefficient is \SI{0.97}{}. \rev{Moreover, we further validated this relationship across an additional user device: a Mac Mini using the same dataset with resolutions 540p, 720p, 1080p, and 2160p. As shown in Fig.~\ref{fig:time_energy2}, decoding time and energy remain strongly correlated, with a Pearson correlation coefficient consistently above \SI{0.96}{}. Although absolute time and energy values differ by platform, the proportional trends confirm that decoding time serves as a robust proxy for decoding energy in practice.} Therefore, decoding time-based estimation provides a more practical and user-centric approach, ensuring efficient playback while indirectly promoting energy efficiency. Moreover, most consumers accept lower video quality (to a certain extent) if it means consuming less energy and saving money~\cite{qoe_energy_tradeoff}.

One significant challenge in adaptive streaming is that some devices support only a fixed set of bitrates. This limitation is often due to hardware constraints or firmware design choices, which only optimize specific bitrates and resolutions for decoding. Frequent jumps in selected resolution for each target bitrate may lead to quality inconsistencies if not designed correctly. Notably, rapid changes in resolution can lead to noticeable quality fluctuations, resulting in a jarring viewing experience. While our optimization is performed offline, targeting video-on-demand (VOD) encoding or initial ladder generation for adaptive streaming, the resulting Pareto-optimal representations can be integrated into real-time adaptive frameworks, such as RL- or MPC-based bitrate selection agents.

This paper builds on our previous work~\cite{KatsenouVCIP2024} and \rev{contributes further to the state of the art with} the following:
\begin{enumerate}
\item \emph{PF optimization under multiple strategies:}
We design and evaluate various ladder construction methods based on multi-objective PF optimization of the rate-quality-time (RQT) parameter space by exploring different strategies, specifically targeting VVC streaming. Unlike prior multi-codec studies~\cite{Ghasempour_Access2024, Ghasempour_ref}, our approach focuses on VVC and provides fine-grained control of trade-offs between rate, perceptual quality, decoding time, and energy consumption.

\item \emph{Monotonic ladder construction with quality assurance:} We introduce a monotonicity constraint that enforces non-decreasing perceptual quality across resolution switches, addressing a key limitation of existing convex-hull-based methods~\cite{Ghasempour_Access2024, Ghasempour_ref}. This ensures perceptual consistency across bitrate ladders and avoids user experience regression from quality dips.

\item \emph{Multi-metric quality-energy tradeoff evaluation:} We move beyond VMAF-centric evaluations and incorporate both VMAF and XPSNR. Our measurement-driven approach directly couples decoding runtime with device-level energy consumption, providing robust and scalable insights into the quality-energy tradeoffs.

\item \emph{Comprehensive content-aware analysis and benchmarking:} We enrich our evaluation by leveraging content-complexity features (via video complexity analyzer (VCA)~\cite{vca_ref}) to interpret how spatio-temporal content affects ladder behavior, and by comparing against state-of-the-art methods under diverse encoding/decoding setups. This results in the first systematic evidence of how content type influences joint RQT–energy optimization for VVC.
\end{enumerate}

\subsubsection*{Paper outline} 
In the following sections, we discuss the state-of-the-art in Section~\ref{sec: SotA_PFStreaming}. Then, we describe our methodology in Section~\ref{sec: Methodology} and present the experimental setup in Section~\ref{sec: Evaluation}. We analyze findings in Section~\ref{sec: Results}. Finally, we summarize our conclusions and outline future research in Section~\ref{sec: Conclusions}.

\section{State-of-the-art in optimized adaptive video streaming}
\label{sec: SotA_PFStreaming}
Two main terms are encountered in adaptive video streaming: PF optimization and convex hull optimization.
Convex hull optimization is stricter than PF as it does not allow non-monotonic solutions. However, in some cases, it may result in a sparse hull, leading to fewer candidate points for the rungs when building the bitrate ladder. 
Netflix~\cite{netflix_paper} performed optimization based on RQ tradeoffs per video or title (for short-duration videos) or on a per-scene (if the video duration is long) basis~\cite{Neparidze, Satti} in their content delivery pipeline. This yielded optimized results as evident from Fig.~\ref{fig:rate_quality}, where the RQ curves across different spatial resolutions are diverse for a set of 250 short video sequences with distinct content, revealing different compression performances and behavior in the parameter space.

\subsection{Ladders based on RQ Optimization}
The first widely adopted solution for streaming video over different network bandwidths was the solution of fixed ladders, \ie a list of fixed combinations of spatiotemporal resolutions for a set of target bitrates for any video content. Fixed ladders, such as those specified in Apple's HLS bitrate ladder~\cite{HLS_ladder_ref}, are not well-suited for all video content. Therefore, early efforts from Netflix~\cite{netflix_paper} and YouTube~\cite{KokaramICIP2018} focused on content-tailored solutions.
Netflix~\cite{netflix_paper} introduced the idea of convex hull optimization, where an extensive exploration of the RQ space was performed to minimize bitrate while maintaining acceptable video quality. To this end, they utilized brute-force encoding to construct the convex hull by encoding videos at multiple resolutions and an extensive set of quantization parameters (QPs). Similarly, Katsenou~\etal~\cite{Katsenou_PCS2019, Katsenou_vmaf_ladder} have also employed encoding in combination with interpolation to achieve the same objective at a reduced number of required encodes. Katsenou also introduced the idea of predicting the crossover points of the RQ curves to optimize ladder prediction further. Adopting this idea, Durbha~\etal~\cite{Krishna_PCS2024} recently presented a similar approach but primarily focused on using visual information fidelity to predict HEVC ladders. Takeuchi~\etal~\cite{Takeuchi_ref} optimized the bitrates and proposed an encoding solution based on Support Vector Regression (SVR), test encodings, and perceptual video quality. They generated multiple bitrate-resolution pairs (encoding recipe), keeping a constant just-noticeable difference~\cite{jnd_streaming, jingwen_vmaf_jnd_sur} gap. 

% \subsection{Advancements in quality and computational considerations}
Recent advances have proposed more dynamic and adaptive techniques incorporating various quality metrics and computational considerations. For example, Nasiri \etal~\cite{ensemble_learning_vvc_ladder} and Katsenou \etal~\cite{Katsenou_IEEEOJSP2021, Katsenou_vmaf_ladder} have explored ensemble methods and machine learning models to predict encoding outcomes and optimize bitrate ladders. These methods aim to minimize bitrate while ensuring high video quality.
Additionally, techniques such as \jtps~\cite{jtps_ref} use random forests to predict encoding configurations, eliminating the need for extensive pre-encodings. This approach, along with others like \vexus~\cite{menon2024convexhull_xpsnr}, employs machine learning models, such as XGBoost, to dynamically optimize encoding configurations.

However, all the aforementioned methods, either as brute-force algorithms or as ML frameworks, target the same RQ optimization from the provider's perspective. The complexity reductions they offer are usually achieved through reduced encoding for constructing the bitrate ladders, which concerns content providers. Hence, they do not consider other important aspects of video streaming, such as the end-user decoding complexity and energy consumption.

\subsection{Incorporating End-user Performance Factors into Optimization}
One significant contribution to PF optimization in video streaming is integrating decoding complexity into the optimization process. Herglotz~\etal~\cite{Herglotz_2019} have addressed the need for efficient playback on resource-constrained devices by incorporating decoding complexity into their optimization frameworks. This ensures that the selected encoding configurations are efficient in terms of bitrate and quality, and feasible for real-time playback on devices with limited computational resources. Azimi~\etal~\cite{Azimi_EUSIPCO2024} proposed to include the decoding time as a constraint to the predicted ladders, ensuring that the decoding time remains below the threshold to mitigate adverse QoE issues such as rebuffering. 
In our previous work~\cite{KatsenouVCIP2024}, we formulated a multi-objective PF optimization problem that balances bitrate and decoding time for VVC streaming.

Moreover, recent studies have explored the potential for optimization in the energy-quality space, rather than the traditional RQ space. Katsenou~\etal~\cite{Katsenou_ICIP2024} have investigated energy-efficient streaming solutions that optimize the tradeoffs between energy consumption and video quality, revealing opportunities for sustainable adaptive video streaming. Rajendran~\etal~\cite{RajendranVCIP2024} and Ghasempour~\etal~\cite{Ghasempour_Access2024, Ghasempour_ref} utilize decoding energy consumption and representation quality, along with a set of bitrates, to construct the energy-aware bitrate ladder. Azimi~\etal~\cite{azimi_qomex} introduced an ML-assisted architecture to select encoding parameters balancing energy consumption and VMAF.

\begin{figure*}[t]
\centering
\begin{subfigure}{\linewidth}
    \centering
    \includegraphics[width=0.975\textwidth]{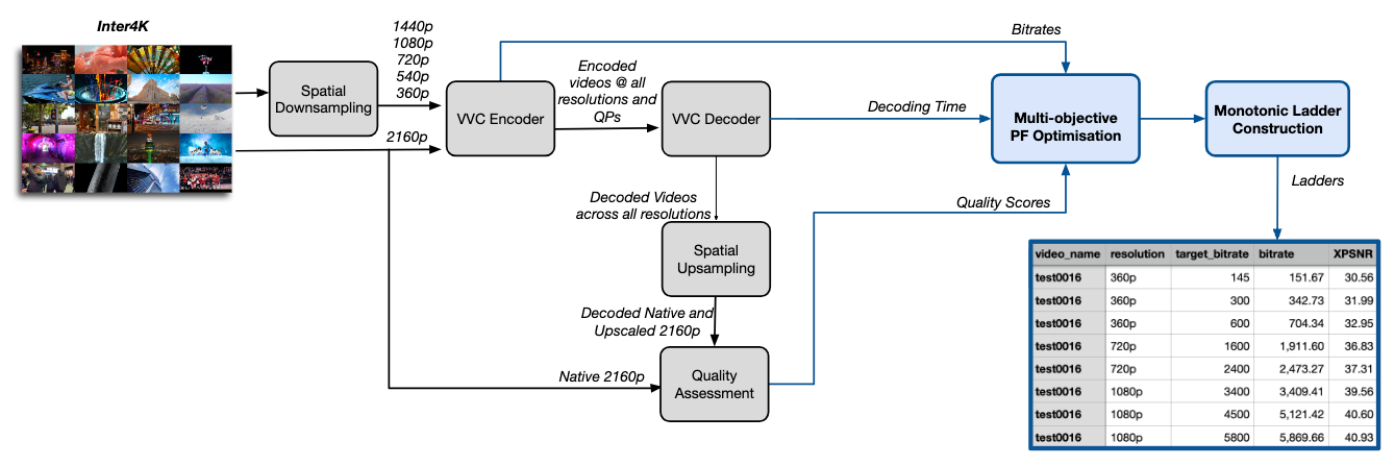}
\end{subfigure}
\caption{Overview of the proposed methodology for the multi-objective PF optimization. Black boxes and lines indicate typical processes and information flow, while blue denotes the particular focus of the proposed method.}
%\vspace{-1.0em}
\label{fig: Overview}
\end{figure*}

\subsection{Moving Beyond the State-of-the-art}
Building on the foundations of the state-of-the-art, including our previous work~\cite{KatsenouVCIP2024},and complementary studies such as~\cite{Ghasempour_Access2024,Ghasempour_ref}, this paper explores the application of a multi-objective PF optimization in VVC streaming. While~\cite{Ghasempour_Access2024,Ghasempour_ref} investigate cross-codec andlive-streaming trade-offs between quality and energy, our work focuses exclusively on VVC and introduces monotonic, measurement-driven Pareto optimization tailored to its higher decoding complexity. Notably, our approach uses decoding time as a proxy for decoding complexity. We evaluate our method and its potential improvements against state-of-the-art methods using a set of key performance indicators, including the quality of the delivered content, decoding time, and resulting energy consumption.

Recent works, such as~\cite{Katsenou_ICIP2024, Ghasempour_Access2024, RajendranVCIP2024} and live streaming variants~\cite{Ghasempour_ref}, propose energy-aware representation selection by incorporating decoding energy and resolution switching into ladder construction. These approaches represent significant progress, but several limitations remain when applied to VVC-based streaming systems. First, many rely on VMAF as the sole quality metric. While VMAF is widely used, recent studies~\cite{wien_xpsnr_vs_vmaf,itu_xpsnr_vs_vmaf} \rev{and the CVQM dataset reports~\cite{cvqm_phase1,cvqm_phase1b} indicate that its correlation with subjective quality for VVC-coded content is not consistently strong. In contrast, XPSNR has shown comparable alignment with human judgments for UHD VVC sequences. In our work, we therefore use XPSNR as a complementary metric. Second, prior methods differ in how they construct representation ladders: some rely on statistical prediction or regression models to reduce encoding overhead (\eg \cite{jtps_ref}), others build convex-hull ladders from measured RQ points (\eg \cite{Katsenou_vmaf_ladder, Menon_ICME2022}). Still, others optimize under energy/time constraints using empirical device measurements (\eg \cite{Azimi_EUSIPCO2024}). Our approach belongs to the last group, but is distinctive in that it performs an entirely empirical, measurement-driven optimization while enforcing quality monotonicity, ensuring reproducibility across devices.} Third, live-streaming adaptations in~\cite{Ghasempour_ref, Azimi_EUSIPCO2024} use regression-based resolution selection to reduce encoding time. While efficient, such strategies do not explicitly enforce monotonic quality consistency. In contrast, our method directly measures decoding time and upscaling cost, enabling stable performance across a diverse range of client devices.

Finally, our work is tailored to the unique characteristics of VVC, addressing its higher computational demands while ensuring perceptual quality stability across the bitrate ladder. By integrating measurement-driven optimization with monotonicity constraints, our approach delivers complexity-aware adaptation tailored to VVC, complementing and extending existing cross-codec and live streaming solutions.

\section{Proposed Multi-objective Pareto-front Optimization}
\label{sec: Methodology}
As illustrated in Fig.~\ref{fig:ParamSpace} for the Inter-4K dataset~\cite{inter4k_ref}, the RQT parameter space across different compression levels and spatial resolutions varies. Higher resolutions and lower QP values enhance video quality but increase bitrate and decoding time. In contrast, lower resolutions and higher QP values reduce decoding time and bitrate at the expense of video quality. The distinct separation of resolutions in the visualizations suggests that adaptive streaming strategies can more precisely customize their bitrate ladders to achieve the desired balance between quality and decoding complexity.

To this end, we propose a methodology that allows us to formulate and solve a multi-objective PF optimization problem. Fig.~\ref{fig: Overview} outlines the workflow for constructing adaptive streaming bitrate ladders by encoding, decoding, and assessing video quality across different resolutions and quantization parameters (QPs). Initially, spatial downsampling is applied to the native $2160\textrm{p}$ content to create lower-resolution versions, $\{1440\textrm{p}, 1080\textrm{p}, 720\textrm{p}, 540\textrm{p}, 360\textrm{p}\}$, which are all encoded along with the native resolution using VVC and an extensive range of QPs. The videos are then decoded using a VVC decoder to generate reconstructed videos at various resolutions and QPs. These decoded streams are further processed through spatial upsampling to upscale lower-resolution videos to $2160\textrm{p}$, allowing quality comparison against the native $2160\textrm{p}$ content. The next phase involves using bitrates, quality scores, and decoding times for each video to populate the parameter space and construct the PF according to multiple strategies (combinations) of the three optimization objectives.
The following step is to sample the PF to construct the bitrate ladders tailored for the video content. The ladder construction imposes monotonicity conditions to support scalable quality of user experience.

In the following subsections, different strategies for combining optimization objectives are presented. 
The rationale behind these strategies lies in scenarios where streaming service providers aim to deliver a consistent viewing experience across a wide range of devices, from high-end smart TVs to budget smartphones, and over a fixed set of target bitrates to ensure a seamlessly switching between representation when the content changes or when the network speed changes, minimizing the risk of playback issues such as buffering, stalls, or quality fluctuations. The proposed methodology is also presented as Algorithm~\ref{algo:rqt_pf}, which includes the required computations and equations for implementing the different multi-objective optimization strategies. 
Our optimization framework assumes that reducing decoding complexity at the encoding stage translates to a proportional decrease in decoding time across all end-user devices. If our optimization strategy reduces decoding complexity by X\%, a similar X\% reduction is expected across various playback devices, including mobile phones, tablets, smart TVs, and desktop computers. While individual device architectures may introduce variations, we assume the relative reductions remain consistent.

\rev{We propose two complementary strategies, \jrqtpf and \jqtpf, that differ in their optimization priorities. \jrqtpf first minimizes rate, then jointly balances quality and decoding time, making it more suitable when bandwidth efficiency is the primary constraint (\eg mobile networks). In contrast, \jqtpf treats quality and time jointly as a unified objective before considering rate, which is advantageous when preserving perceptual consistency and smooth playback is paramount (\eg premium UHD streaming or battery-constrained devices). While the resulting Pareto fronts can appear similar, the distinction in formulation allows service providers to adopt the strategy best aligned with their operating requirements and constraints.}

\begin{algorithm}[t]
\caption{Multi-objective PF Optimization}
\small
\textbf{Input:} set of supported resolutions, $\mathcal{R}$, and set of QPs, $\mathcal{Q}$\\
\begin{algorithmic}[1]
\STATE Load video content \;
\FOR{each $r \in \mathcal{R}$}
    \FOR{each $q \in \mathcal{Q}$}
        \STATE Encode video using ($r, q$) \;
        \STATE Compute the bitrate ($b$) \;
        \STATE Measure decoding time ($\tau_{D}$) \;
        \STATE Calculate quality $v$ (\eg XPSNR, VMAF) \;
    \ENDFOR
\ENDFOR
\STATE Calculate the optimization objectives; $M$ as in Eq.(\ref{eq: M}) for \jrqtpf or $J$ as in Eq.(\ref{eq: J}) for \jqtpf\;
\STATE Perform multi-objective optimization in the $M-v$ or $J-b$ space to identify PF points as in Eq.(\ref{eq: Mopt})\;
\STATE Construct monotonic bitrate ladders as in Eq.~(\ref{eq: monotonicQuality}). \;
\end{algorithmic}
\vspace{-.5em}
\label{algo:rqt_pf}
\end{algorithm}

\subsection{Joint Quality-Time PF (\jqtpf) strategy}
The Joint Quality-Time PF (\jqtpf) is designed to optimize video quality and decoding time while adhering to a fixed set of bitrates. This optimization is essential because it balances the need for high video quality with the practical constraints of decoding performance. It is vital for devices with limited computational power or energy resources.

To formalize this optimization, we define a composite utility function $J$ that combines two antagonists: video quality $v$ and decoding time $\tau_{D}$, where effectively high quality is penalized for long decoding times. This utility function is formulated as follows:
\begin{equation}
J = v-\alpha_J \log(\tau_{D}) 
\label{eq: J}
\end{equation}
where $\alpha_J \in \mathbb{R}^+$.
After defining this composite utility function, the goal is to maximize $J$ while minimizing the bitrate $b$. This can be formulated as follows:

\begin{equation}
  \begin{split}
    \max_{r, q} J  & =  \max_{r, q} \; v-\alpha_J \log(\tau_{D})\\
    \min_{r, q} b&  \\
    \text{subject to: } &
    \begin{aligned}[t]
      r \in \mathcal{R} & \\
          q\in\mathcal{Q} &  
    \end{aligned}
  \end{split}
  \label{eq: Jopt}
  \end{equation}
This setup allows for direct computation of the $J-b$ PF, ensuring a thorough exploration of trade-offs. In these equations, $\alpha_J$ is the weight assigned to the logarithm of the decoding time $\tau_{D}$. The logarithmic transformation of the decoding time is used to appropriately scale its wide range of values, ensuring that the metric remains sensitive to high and low decoding times. The video quality metric $v$ can be measured using metrics such as XPSNR~\cite{XPSNR} or VMAF~\cite{VMAF}. \vig{The optimization searches over all valid resolution and QP combinations $(r, q)$ in the sets $\mathcal{R}$ and $\mathcal{Q}$, respectively. We compute the corresponding $(v, \tau_D, b)$ triplet for each candidate representation and evaluate its objective score $J$. Among all feasible configurations, we select those that maximize $J$ while ensuring quality monotonicity across the ladder (\ie quality should not decrease when moving to higher bitrates). Although bitrate $b$ is not directly included in the objective, it is minimized implicitly by preferring representations with lower $b$ when multiple candidates have similar $J$ values.}
 
The \jqtpf consists of a set of encoding configurations for which it is impossible to improve one objective without degrading the other. In this strategy, mathematically, a point ($J, b$) is on the PF iff there is no other point ($J', b'$) such that:
\begin{align}
J' \geq J \land b' \leq b
\end{align}
with at least one strict inequality.

\subsection{Joint Rate-Quality-Time PF (\jrqtpf) strategy}
This multi-objective \jrqtpf strategy was proposed in our previous work~\cite{KatsenouVCIP2024}. 
Based on this strategy, the bandwidth efficiency through bitrate and the decoding complexity through decoding time must be optimized simultaneously. 
Towards this realization, we define a composite metric $M$ as a linear combination of decoding time and bitrate, weighted by parameter $\alpha_M$:
\begin{align}
    M = \alpha_M \cdot log (\tau_{D}) + (1-\alpha_M)  \cdot log (b),
    \label{eq: M}
\end{align}
As in the previous strategy, the logarithmic transformation of decoding time and bitrate allows for the appropriate scaling of their wide range of values. 
The goal is to minimize the composite metric $M$ while maximizing video quality $v$. This can be formulated as a multi-objective optimization problem:
\begin{equation}
    \begin{aligned}
        \min_{r,q} M & = \min_{r, q} (\alpha_M \log_{10}(\tau_{\text{D}}) + (1-\alpha_M) \log_{10}(b))& \\
        \max_{r, q} v & & \\
% \text{with}& \alpha \in [0, 1] &\\
% % &\alpha + \beta = 1 &\\
%  &
    \end{aligned}
\label{eq: Mopt}
\end{equation}
with $\alpha_M \in [0, 1]$.
\vig{To implement this, we generate the complete set of candidate representations by encoding each video sequence at all combinations of ($r, q$). We compute $b$, $\tau_D$, and $v$ for each configuration. $M$ is then calculated using the weighted sum of $\log_{10}(\tau_D)$ and $\log_{10}(b)$, allowing control over the tradeoff between decoding complexity and bandwidth cost via the parameter $\alpha_M$.} Streaming services can leverage these PFs to select optimal encoding configurations that align with their specific constraints and user needs. For example, a service prioritizing low latency might opt for configurations from the PF with $\alpha_M = 0.75$, whereas one focusing on bandwidth efficiency would likely choose $\alpha_M = 0.25$. When $\alpha_M = 1$, the primary goal shifts to minimizing decoding time relative to bitrate while ensuring high quality, resulting in a PF derived solely from Quality-Decoding Time curves across spatial resolutions. Conversely, an $\alpha_M = 0$ value disregards decoding time, constructing the PF exclusively from RQ curves consistent with standard practices (\eg \cite{netflix_paper}). This approach will henceforth be referred to as \texttt{DynResXPSNR}, as introduced in~\cite{menon2024convexhull_xpsnr}.

Similarly to the previous strategy, here, a point ($M, v$) is on the \jrqtpf, if and only if there is no other point ($M', v'$) such that:
\begin{align}
M' \leq M \land v' \geq v
\end{align}
with at least one strict inequality.

The \jqtpf and \jrqtpf strategies have a similar logic of reducing the dimensionality of the optimization problem by combining two of the three target objectives: rate, quality, and decoding time. Consequently, the resulting solutions are expected to exhibit similar behaviors with some trade-off variations. 

\subsection{Constructing Monotonic Ladders}
The initial step in constructing the bitrate ladder selects the target bitrates that will represent the rungs, \ie $\mathcal{B}={b_{L,1},b_{L,2},\ldots,b_{L,|\mathcal{L}|}}$, where $|\mathcal{L}|$ is the cardinality of $\mathcal{B}$ and $b_{L,1}<b_{L,2}<\ldots<b_{L,|\mathcal{L}|}$. The bitrate ladder is fully defined as a set of tuples $\mathcal{L}$ that comprise bitrate values $\mathcal{B}$, the associated set of quality values $\mathcal{V}$, a set of QP values $\mathcal{Q}$, and a set of resolutions $\mathcal{R}$, \ie
\begin{equation} \label{eq: Ladder} 
\mathcal{L}:= {\langle b_{L,i},v_{L,i},q_{L,i},r_{L,i} \rangle }^{|\mathcal{L}|}, 
\end{equation} 
\noindent
with $b_{L,1}<b_{L,2}<\ldots<b_{L,|\mathcal{L}|}$. 

A critical factor in constructing a bitrate ladder is ensuring the monotonicity of the resulting rungs. Although rate monotonicity is inherently imposed by the ranked sampling intervals above, monotonicity in the quality dimension is not guaranteed, particularly when exploring multiple dimensions of the parameter space. This lack of monotonicity in quality can pose challenges for video providers, as it may degrade the end-user's quality of experience. To address this issue, we recommend enforcing monotonicity by ensuring consistently increasing monotonic quality (\eg XPSNR) across rungs, \ie
\begin{equation} 
\label{eq: monotonicQuality} 
    v_{L,1}\leq v_{L,2}\leq \ldots \leq v_{L,|\mathcal{L}|} \, .
\end{equation}
In prior research~\cite{Katsenou_IEEEOJSP2021, menon2024convexhull_xpsnr, Krishna_PCS2024}, where the parameter space considered was only the RQ space, a different approach was followed to reinforce monotonicity in the constructed ladders. Monotonicity was reinforced through the adoption of increasing monotonic resolution switching across rungs $r_{L,1}\leq r_{L,2}\leq \ldots \leq r_{L,|\mathcal{L}|}$. However, in the multi-objective PF optimizations proposed in this paper, frequent non-monotonic resolution switching enables the selection of representations with similar or better video quality within the target bitrate range, potentially using lower resolutions that offer reduced decoding times. Therefore, we only impose video quality monotonicity.

\section{Evaluation Setup}
\label{sec: Evaluation}
This section illustrates our experimental design to assess the performance of the multi-objective PF optimization strategies. We compare our strategies against distinct benchmark schemes by establishing performance metrics for coding performance, runtime, and energy consumption.

\subsection{Benchmarks}
The proposed method is compared against the following state-of-the-art methods:
\noindent
\renewcommand{\labelenumi}{\alph{enumi}.}
\begin{enumerate}[leftmargin=*]
    \item \fixedladder is based on the HLS bitrate ladder as a fixed set of bitrate-resolution pairs~\cite{HLS_ladder_ref}. This solution only seeks the best QP of a specific spatial resolution for every rung to hit the target rate. 
    \item \default ladder is the simplest case, with no resolution switching. All representations are encoded at 2160p for a given set of bitrates. This method demonstrates a limitation in covering the lower range of the bitrate ladder.
    \item \DynResVMAF selects the resolution and QP yielding the highest VMAF for a given set of bitrates~\cite{netflix_paper}. This is content-gnostic, tailored to each sequence's unique Rate-VMAF curves across resolutions.
    \item \DynResXPSNR is similar to \DynResVMAF, but based on XPSNR instead of VMAF~\cite{menon2024convexhull_xpsnr}.
    \item Azimi~\etal selects the resolution that produces the highest XPSNR for a given set of bitrates with constrains over a maximum decoding time threshold~\cite{Azimi_EUSIPCO2024}. Here, we experiment with $\tau_{L} =\{12, 16, 24, 32, 64\}$ seconds.
    \item \emph{Non-Monotonic JRQT-PF}~\cite{KatsenouVCIP2024} is our previous multi-objective solution that does not impose any quality monotonicity constraints.
\end{enumerate}

\subsection{Dataset}
We used the Inter-4K dataset~\cite{inter4k_ref}, which comprises 1,000 UHD video clips collected from YouTube. Spatial downscaling is applied using FFmpeg’s bicubic filter to generate lower-resolution versions of each clip. \rev{However, the proposed optimization is \emph{kernel-agnostic}; alternative filters (e.g., Lanczos) can be plugged in without changing the pipeline.} The content spans six main categories: urban environments (\eg buildings, streets, and traffic), natural scenes and animals, sports and human activities, visual demos and abstract graphics (including frame rate/resolution demos and synthetic shapes), music videos, and cinematic content. 

\rev{Although our evaluation focuses exclusively on UHD content, this choice was deliberate: UHD streams pose a significant challenge in terms of decoding time and energy consumption, and are therefore a representative stress test for complexity-aware ladder design. Nevertheless, the proposed framework is resolution-agnostic. In practice, the same Pareto-front optimization procedure can be applied to any target resolution. While the absolute values of decoding time and energy naturally decrease with lower resolutions, the relative trade-offs between rate, quality, and complexity remain consistent. Thus, our method scales across different operating resolutions, with UHD results representing the upper bound of computational demand in real-world deployments.}

\subsection{Evaluation Metrics}
\label{ssec: EvalMetrics}
We evaluate the methods using the following metrics:
\begin{description}[font=\normalfont\itshape]
\item[Quality metrics] We compare the overall quality in PSNR~\cite{psnr_ref}, VMAF~\cite{VMAF}, and XPSNR~\cite{XPSNR} for every target bitrate of each test sequence.
% \subsubsection{\rev{Quality–Metric Computation Cost (Offline Analysis)}}
% \label{sec:metric_cost}
\rev{In large-scale ladder construction, the time to compute perceptual metrics can dominate offline evaluation cost. Figure~\ref{fig:metric_times} reports wall-clock times to compute PSNR, XPSNR, and VMAF for 240 UHD frames on our test platform.\footnote{Same workstation configuration as used for decoding-time experiments in Section~\ref{sec:dec_energy_measure}.}
We observe that XPSNR executes much faster than VMAF. In practice, this efficiency enables broader operating-point sweeps without prohibitive evaluation time.}

\begin{figure}[t]
    \centering
    \includegraphics[width=0.16\columnwidth]{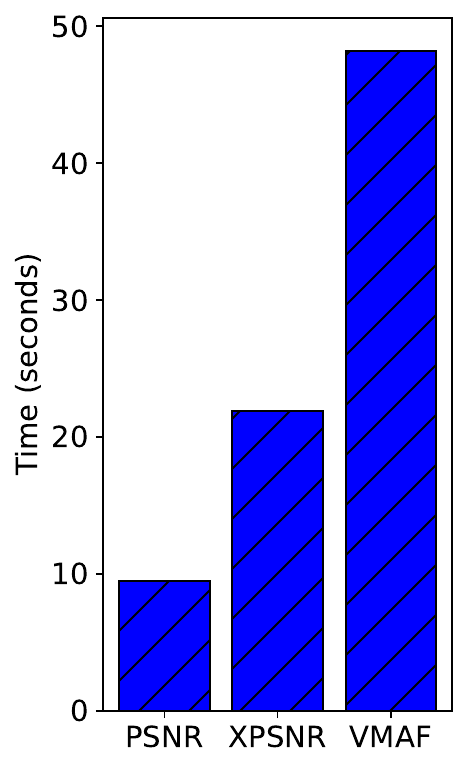}
    \caption{\rev{Wall-clock time to compute PSNR, XPSNR, and VMAF for 240 UHD frames on our platform. XPSNR is substantially faster than VMAF, reducing offline evaluation cost.}}
    \label{fig:metric_times}
\end{figure}

\item[Decoding runtime difference]
$\Delta T_{\text{D}}$
\begin{align}
 \Delta T_{\text{D}} &= \frac{\sum \tau_{method} - \sum \tau_{ref}}{\sum \tau_{ref}}
  % \Delta E_{\text{D}} &= \frac{\sum E_{method} - \sum E_{ref}}{\sum E_{ref}}
\end{align}
\vig{It is important to note that the reported decoding times in this paper include the time required for upscaling each decoded frame to 4K resolution, as used during quality evaluation. This ensures consistency between decoding complexity and perceptual quality assessment, simulating realistic playback conditions on UHD displays. \rev{The time taken for bicubic interpolation was less than 1 ms per frame, which is negligible compared to the times required for VVC decoding.} Moreover, the reported decoding time refers to \emph{user time}, which measures the CPU time consumed by the process in user mode, excluding time spent in system-level operations or waiting for I/O. User time is particularly relevant in our context, as it closely reflects the computational workload handled by the decoder and correlates well with energy consumption and device-side performance in VVC playback scenarios.}
\item[Bjøntegaard Delta metrics] are computed over Rate-PSNR, Rate-XPSNR, and Rate-VMAF ladders. Particularly, rates \BDRP, \BDRX, and \BDRV refer to the average increase in bitrate of the considered methods compared with the \fixedladder encoding with the same \mbox{PSNR}, \mbox{XPSNR}, and \mbox{VMAF}, respectively~\cite{BJDelta}. Moreover, we also evaluate Bjøntegaard Delta Decoding Energy (\mbox{BDDE})~\cite{Herglotz_2019}, which measures energy savings as a percentage for the same video quality. A negative \mbox{BDDE} indicates
that the same quality is achieved with reduced energy consumption.

\end{description}

\begin{table}[t]
\caption{Experimental parameters and values.}
\centering
\resizebox{0.495\columnwidth}{!}{
\begin{tabular}{l||c|c|c|c|c|c}
\specialrule{.12em}{.05em}{.05em}
\specialrule{.12em}{.05em}{.05em}
\emph{Parameter} & \multicolumn{6}{c}{\emph{Values}}\\
\specialrule{.12em}{.05em}{.05em}
\specialrule{.12em}{.05em}{.05em}
$\mathcal{R}$ & \multicolumn{6}{c}{\{ 360, 540, 720, 1080, 1440, 2160 \} } \\
\hline
$\mathcal{B}$ [in Mbps] & 0.145 & 0.300 & 0.600 & 0.900 & 1.600 & 2.400 \\
\cmidrule{2-7}
              & 3.400 & 4.500 & 5.800 & 8.100 & 11.600 & 16.800 \\
\hline
$\mathcal{Q}$ &  \multicolumn{6}{c}{10, 12, \ldots, 50} \\
% \hline
% $\alpha$ & \multicolumn{2}{c|}{0.25} & \multicolumn{2}{c|}{0.5} & \multicolumn{2}{c}{0.75}\\
\hline
Target encoder & \multicolumn{6}{c}{VVenC [medium][intra period=\SI{1}{\second}][4 CPU threads]} \\
\hline
Target decoder & \multicolumn{6}{c}{VVdeC [4 CPU threads]} \\
\specialrule{.12em}{.05em}{.05em}
\specialrule{.12em}{.05em}{.05em}
\end{tabular}
}
%\vspace{-0.98em}
\label{tab:exp_par}
\end{table}

\subsection{Experimental Parameters and Values}
The experimental parameters used in this paper are reported in Table~\ref{tab:exp_par}. We run all experiments on a dual-processor server with Intel Xeon Gold 5218R (80 cores, frequency at 2.10 GHz), where each VVenC encoding instance uses four CPU threads. We measured the CPU energy consumption on Linux using the Running Average Power Limit (RAPL) interface and the \texttt{CodeCarbon} tool~\cite{codecarbon_ref}.
\section{Experimental Results}
\label{sec: Results}
Based on the evaluation design in Section~\ref{sec: Evaluation}, we structure this section based on three main aspects: coding efficiency, decoding runtime and energy consumption.

\begin{figure}[t]
    \centering
    \includegraphics[width=0.45\textwidth]{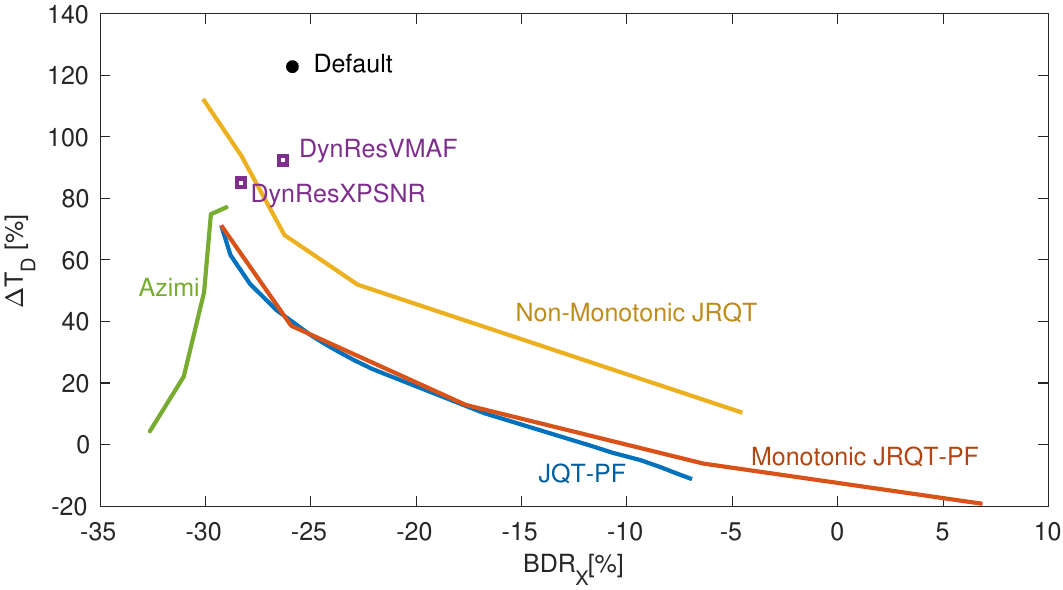}
    \caption{Comparison of \BDRX-$\Delta T_{\text{D}}$ tradeoffs of all compared methods. Ideally, negative values for both dimensions would give the best tradeoffs compared to \fixedladder.
    }
    %\vspace{-2em}
    \label{fig:JRQTvsJQT}
\end{figure}
\begin{table*}[t]
\caption{Average coding performance of the explored and proposed methods for selected weights compared against \fixedladder. Arrows indicate good performance trends. }
\centering
\resizebox{0.655\linewidth}{!}{
 
\begin{tabular}{@{}l@{ }|@{}l@{ }||@{ }c@{ }|@{ }c@{ }|@{ }c@{ }|@{ }c@{ }|@{ }c@{ }|@{ }c@{ }|@{ }c@{ }}
\specialrule{.12em}{.05em}{.05em}
\specialrule{.12em}{.05em}{.05em}
Method & Parameter & \BDRP  & \BDRX  & \BDRV & BD-PSNR  & BD-XPSNR & BD-VMAF & $\Delta T_{\text{D}}$  \\
& & [\%] $\downarrow$ & [\%] $\downarrow$ & [\%] $\downarrow$ & [dB] $\uparrow$ & [dB]$\uparrow$ & $\uparrow$ & [\%] $\downarrow$  \\
\specialrule{.12em}{.05em}{.05em}
\specialrule{.12em}{.05em}{.05em}
\default & - & -30.12 &	-25.84 &	-44.72 &	1.63 &	1.37 &	8.79 &	122.77  \\
\hline
\DynResVMAF~\cite{netflix_paper} & - & -31.64 &	-26.31 & -50.28 &	1.57 &	1.27 &	10.51   & 92.34 \\
\DynResXPSNR~\cite{menon2024convexhull_xpsnr} & - & -32.29 & -28.32 & -48.99 & 1.59 & 1.34 & 10.22 & 85.07 \\
\hline
\multirow{2}{*}{\emph{Azimi}~\etal~\cite{Azimi_EUSIPCO2024}} &
$\tau_{L}=$ \SI{12}{\second} & -38.81 & -32.64 & -53.86 & 1.59 & 1.27 & 13.81 & 4.31 \\
           
 & $\tau_{L}=$ \SI{64}{\second} & -31.73 &	-29.00 &	-47.14 &	1.48 &	1.30 &	9.72 &	77.10 \\
 \hline
\multirow{5}{*}{\emph{JRQT-PF}$^{*}$~\cite{KatsenouVCIP2024}} & $\alpha_M=1$ & -4.62 &	-4.52 &	5.55 &	0.42 &	0.39 &	0.84 &	10.35  \\
 & $\alpha_M=0.75$ & -24.44 & -22.77 &	-6.05 &	1.18 &	1.07 &	6.04  &	51.93  \\
 & $\alpha_M=0.5$ &  -28.29 &	-26.25 & -38.39 &	1.36 &	1.23 &	7.58 &	68.06  \\
 & $\alpha_M=0.25$ & -31.25 &	-28.28 & 	-45.76 &	1.53 &	1.34 &	9.27 &	93.67  \\
 & $\alpha_M=0$ & -32.90	& -30.11 &	-47.54 &1.63 &	1.43 &	9.98 &	112.26   \\
 \hline
\multirow{5}{*}{\emph{JRQT-PF}$^{\dagger}$} & $\alpha_M=1$ &   7.40&	 6.87 &	 32.88&	 -0.03&	 -0.04&	-1.61 &	-19.24\\
 & $\alpha_M=0.75$ &  -6.58 & -6.38 & 0.66 & 0.45 & 0.41 & 1.07 &  -6.17  \\
 & $\alpha_M=0.5$ &  -18.62 & -17.64 & -18.92 & 0.92 & 0.84 & 4.04 &  12.86	  \\
 & $\alpha_M=0.25$ &   -27.57 & -25.93 & -34.90 & 1.33 & 1.20 & 7.07 & 38.60  \\
 & $\alpha_M=0$ & -31.95 & -29.25 & -46.11 & 1.54 & 1.35 & 9.36 &  71.19 \\
\hline
\multirow{5}{*}{\emph{JQT-PF}$^{\dagger}$} & $\alpha_J = 3$ & -7.11 & -6.91 & 10.40 & 0.53 & 0.49 & 0.99 & -11.25 \\
& $\alpha_J = 2.5$ & -12.04 & -11.76 & -3.48 & 0.71 & 0.66 & 2.15 & -0.29\\
& $\alpha_J = 2.0$ & -17.53 & -16.78 & -12.89 & 0.92 & 0.85 & 3.59 & 10.22\\
& $\alpha_J = 1.5$ & -22.34 & -21.22 & -23.40 & 1.11 & 1.02 & 5.02 & 22.21\\
& $\alpha_J = 0.5$ & -29.88 & -27.88 & -39.45 & 1.44 & 1.29 & 8.01 & 52.16\\
\specialrule{.12em}{.05em}{.05em}
\specialrule{.12em}{.05em}{.05em}
\end{tabular}
}
\begin{flushright}
\footnotesize * represents non-monotonic, and $^{\dagger}$ represents XPSNR-monotonic.
\end{flushright}
%\vspace{-1em}
\label{tab:res_cons_energy}
\end{table*}

\subsection{Coding Efficiency Analysis}
We experimented with the different $\alpha$ values for the multi-objective strategies to assess their impact on the generated ladders. The results were evaluated primarily using the Bjøntegaard Delta (BD) metric for rate and quality, comparing the average encoding performance of various encoding methods against \fixedladder. These results are summarized in Fig.~\ref{fig:JRQTvsJQT}, and some representative results are included in Table~\ref{tab:res_cons_energy}. 

\begin{figure}[t]
\centering
\begin{subfigure}{0.34\textwidth}
     \centering
     \includegraphics[width=1\textwidth]{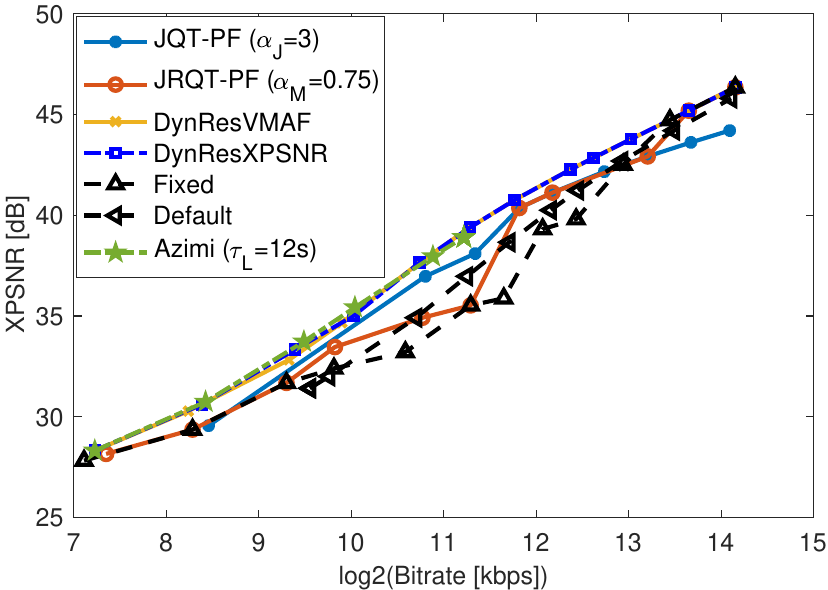}
      \includegraphics[width=1\textwidth]{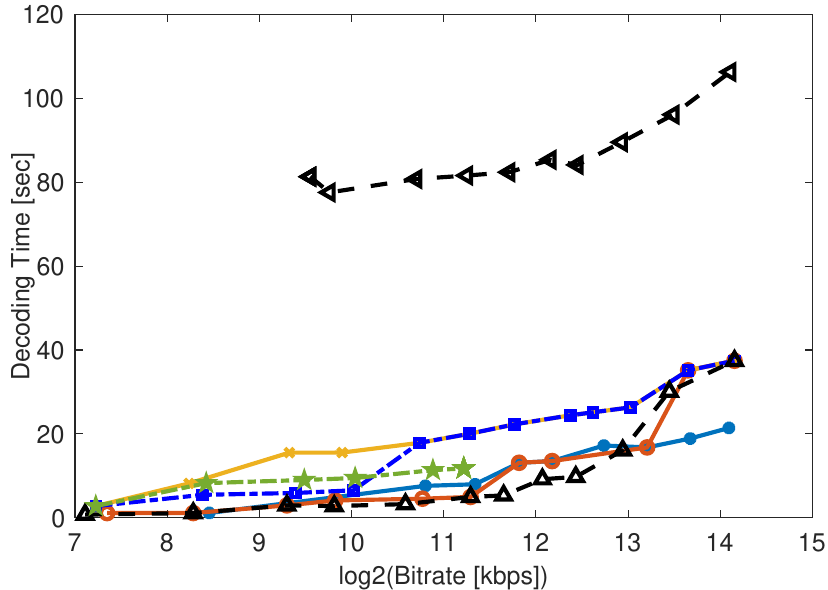}
\caption{Sequence \textit{0263}\\ (\EY:125.54, \h:34.81, \LY: 116.45)}    
\end{subfigure}
\begin{subfigure}{0.34\textwidth}
     \centering
     \includegraphics[width=1\textwidth]{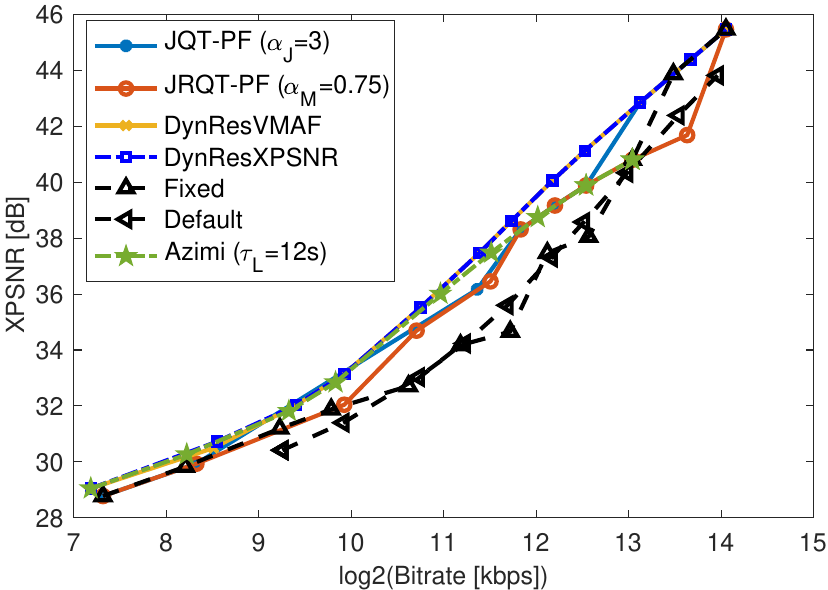}
     \includegraphics[width=1\textwidth]{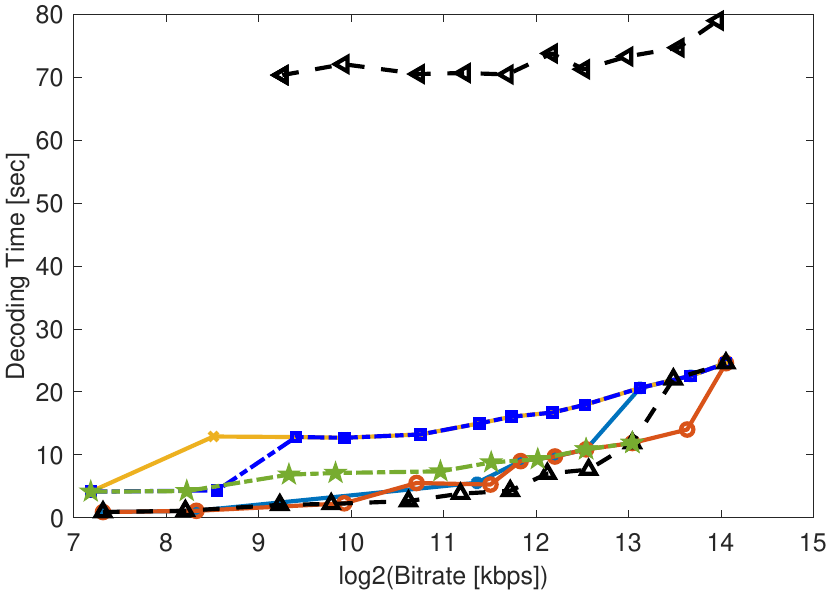}
\caption{Sequence \textit{0276}\\ (\EY:191.49, \h:15.60, \LY:123.40)}     
\end{subfigure}

\caption{\vig{Resulting ladders of four example video sequences using selected benchmark methods, \jrqtpf for $\alpha_M=0.75$, and \jqtpf for $\alpha_J=3$.}} 
%\vspace{-1.5em}
\label{fig:rd_res}
\end{figure}
Based on those, \default shows significant bitrate savings and the highest quality gains in terms of PSNR, for the bitrate range it can cover across all BD metrics (lower bitrates are excluded). \DynResXPSNR and \DynResVMAF perform even better in terms of bitrate reduction and quality improvement. As shown in Fig.~\ref{fig:JRQTvsJQT}, all multi-objective strategies benefit from a regulated compression performance. Using different weight values, the gains in rate and quality can be enhanced or reduced. The monotonic version of \jrqtpf shows improvements in rate savings and gains in quality compared to the non-monotonic version. It is also easy to notice that \jqtpf performs similarly to the monotonic \jrqtpf but with the advantage of better regulation of the tradeoffs.  Ideally, in this figure, negative values for both dimensions would give the best tradeoffs compared to \fixedladder. Although the Azimi~\cite{Azimi_EUSIPCO2024} curve for different thresholds, $\tau_L = \{12, 16, 24, 32, 64\}$ moves to the right direction, the \jqtpf (for $\alpha_J\geq2.5$) method appears to be offering the best $BDR_X \textrm{-} \Delta T_D$ tradeoff.

For better comprehension, we also provide a visual example of the ladders produced by the proposed \jqtpf method, along with a selection of benchmarks, in Fig.~\ref{fig:rd_res}. Two sequences were selected based on their different spatio-temporal features extracted using the Video Complexity Analyzer (VCA) tool~\cite{vca_ref}. 
The scene in \emph{Sequence 0263} has
different content features: high spatial texture energy (\EY) and temporal texture energy (\h), but moderate average luminance (\LY). It depicts a crowd of tourists before a
sculpted building with reflections on its window muntins.
The scene in \emph{Sequence 0276} captures a man performing capoeira in a green field with trees under the blue sky, resulting in very high texture energy, low temporal texture energy, and moderate average luminance. The constructed ladders of \DynResVMAF and \DynResXPSNR are almost identical, indicating that they are the best in terms of quality. Azimi~\etal~ strictly limits bitrate choices based on decoding time; hence, there is a risk of unnecessarily capping the video quality. This would also mean that the produced ladders might not cover the entire range of bitrates, as shown in these examples. Thus, this capping may result in a suboptimal viewing experience, particularly for users with high-performance devices. \jqtpf and \jrqtpf follow similar patterns and interlace due to different resolution switches at a subset of the bitrate rungs. It is observed that the sequences with rich spatial textures or fast temporal motion typically yield higher bitrates and decoding times. At the same time, simpler or low-motion content is encoded and decoded more efficiently. These differences validate the importance of content-aware optimization in our framework. \rev{While metric choice does not affect playback-time complexity, it impacts offline evaluation scalability; we therefore report results with both VMAF and XPSNR and favor XPSNR for large sweeps due to its markedly lower computation time (Fig.~\ref{fig:metric_times}).}

\subsection{Decoding Runtime Analysis}
We conduct a comprehensive evaluation of ladder construction schemes by analyzing their relative differences in runtime during decoding ($\Delta T_{\text{D}}$) compared to \fixedladder. The results are reported in Fig.~\ref{fig:JRQTvsJQT} and Table~\ref{tab:res_cons_energy}. 
While \default, \DynResVMAF, and \DynResXPSNR achieve notable bitrate savings, they incur a substantial cost in decoding time, with increases from \SI{74.85}{\percent} to \SI{122.77}{\percent}. On the other hand, Azimi~\etal~\cite{Azimi_EUSIPCO2024}, along with the multi-objective strategies that focus on optimizing quality, rate, and decoding time, demonstrate the potential to balance the decoding time cost for comparable compression performance. For Azimi~\etal, lowering the $\tau_L$ threshold offers significant BD rate gains for a small increase in decoding time compared to \fixedladder. For the multi-objective strategies, significant improvements in $\Delta T_{\text{D}}$ can be achieved for higher $\alpha$ values ($\alpha_M \geq 0.7$ for \jrqtpf and $\alpha_J \geq 2.5$ for \jqtpf) at reduced BD rate gains and quality improvement compared to \fixedladder. Fig.~\ref{fig:JRQTvsJQT} demonstrates the flexibility and effectiveness of the multi-objective strategies in balancing between bitrate, quality, and decoding time according to the selected $\alpha$. Ideally, the best strategies are those that provide BD rate gains ($< 0$) and decoding runtime reductions ($\Delta T_{\text{D}} < 0$). Generally, on average, \jqtpf exhibits the best trade-off compared to other methods. \jqtpf offers the benefit of slower convergence and unbound $\alpha_J$ values (compared to $\alpha_M\in[0,1]$) that could be fine-tuned for better trade-offs.

\begin{figure}[t]
\centering
\begin{subfigure}{0.325\linewidth}
    \centering
    \includegraphics[width=\textwidth]{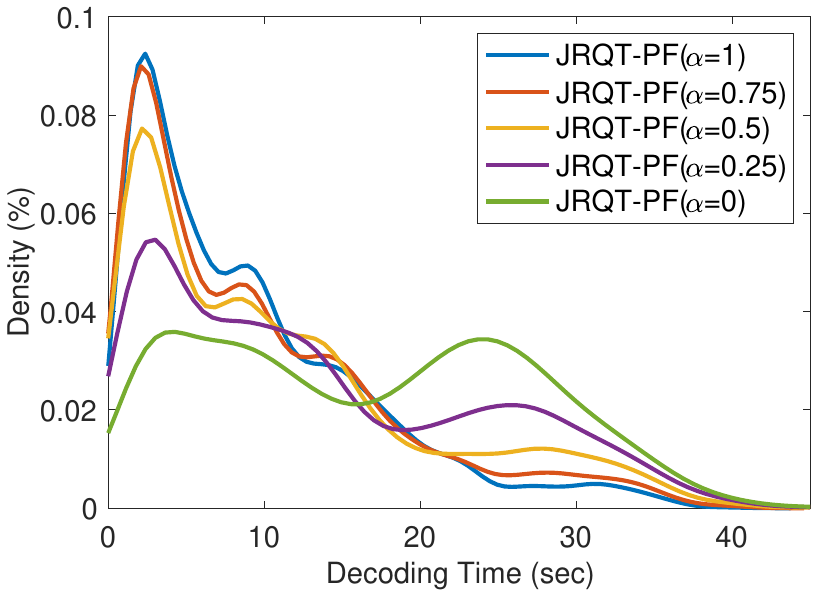}
    \caption{\jrqtpf Decoding Time.}
    \label{fig: JRQTpdf_dec_time}
\end{subfigure}
\begin{subfigure}{0.325\linewidth}
    \centering
    \includegraphics[width=\textwidth]{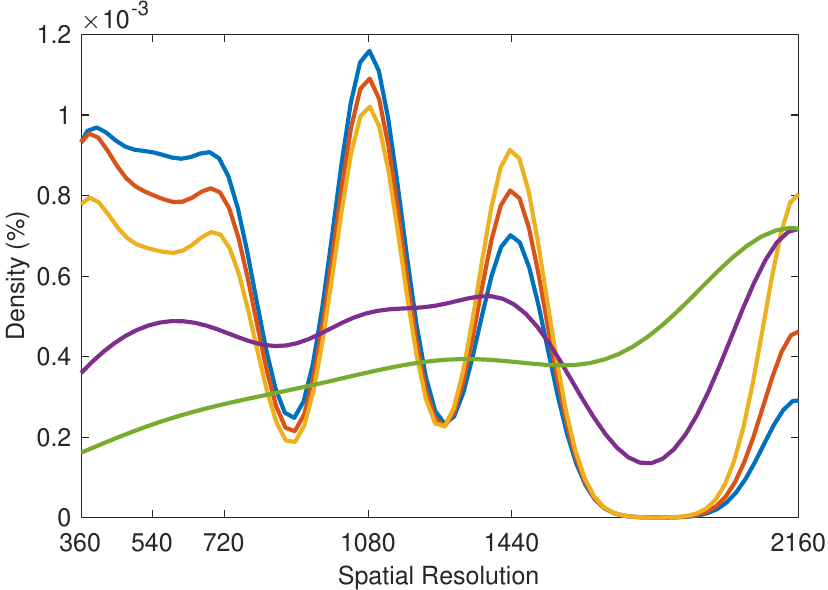}
    \caption{\jrqtpf Spatial Resolution.}
    \label{fig: JRQTpdf_bitrate}
\end{subfigure}
\begin{subfigure}{0.325\linewidth}
    \centering
    \includegraphics[width=\textwidth]{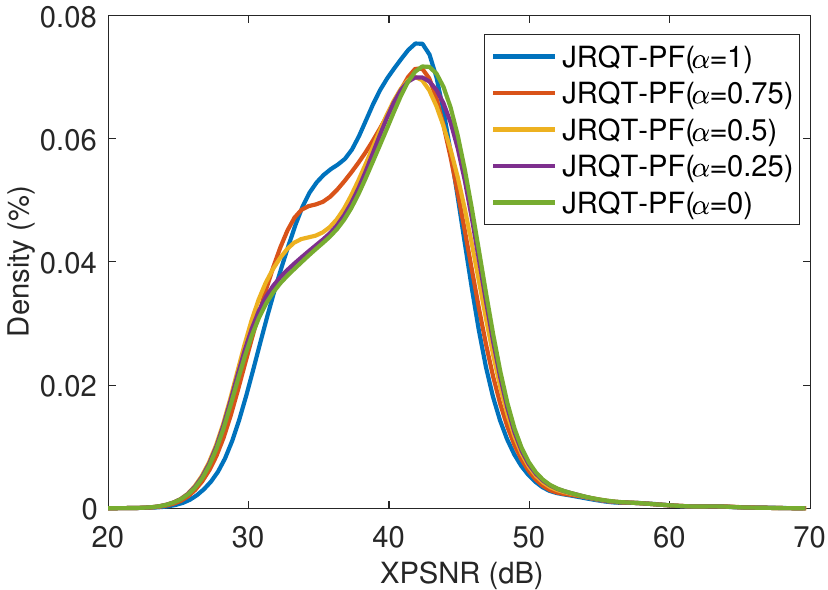}
    \caption{\jrqtpf XPSNR.}
    \label{fig: JRQTpdf_xpsnr}
\end{subfigure}

\begin{subfigure}{0.325\linewidth}
    \centering
    \includegraphics[width=\textwidth]{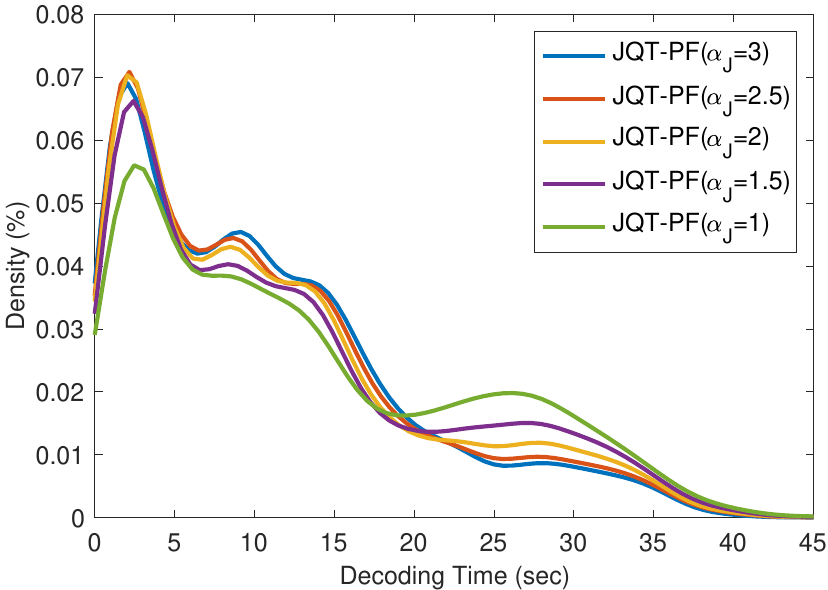}
    \caption{\jqtpf Decoding Time.}
    \label{fig: JQTpdf_dec_time}
\end{subfigure}
\begin{subfigure}{0.325\linewidth}
    \centering
    \includegraphics[width=\textwidth]{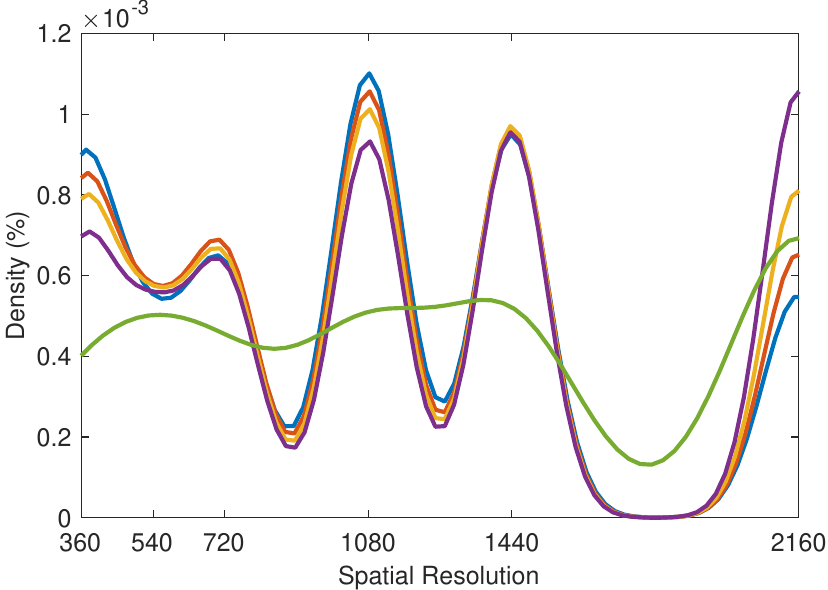}
    \caption{\jqtpf Spatial Resolution.}
    \label{fig: JQTpdf_bitrate}
\end{subfigure}
\begin{subfigure}{0.325\linewidth}
    \centering
    \includegraphics[width=\textwidth]{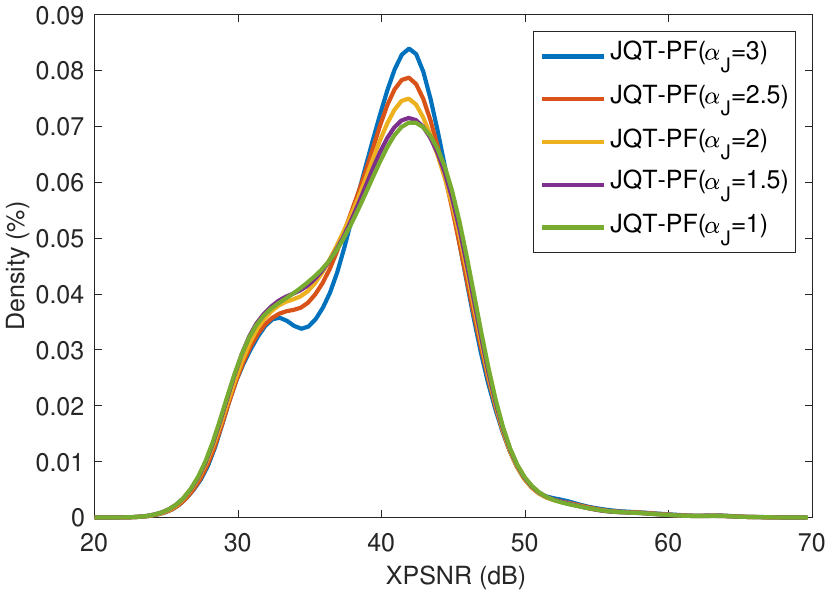}
    \caption{\jqtpf XPSNR.}
    \label{fig: JQTpdf_xpsnr}
\end{subfigure}
\caption{Probability Density Functions (PDFs) for the \jrqtpf Ladders (a, c, e) and \jqtpf (b, d, f).}
%\vspace{-1.5em}
\label{fig: PDFs}
\end{figure}

These findings can be confirmed by the probability density function (PDF) curves of Fig.~\ref{fig: PDFs}. The PDF curve transitions depict the influence of the $\alpha$ values. Varying $\alpha$ values demonstrates the flexibility of the joint optimization approach, achieving a range of decoding times that generally decrease as $\alpha$ increases (with higher peaks for decoding times lower than \SI{5}{\second}). Similar patterns are observed for XPSNR PDFs increase peak values. These shifts are achieved by switching to low-resolution representations, as confirmed by the PDFs. \rev{Note that Fig.~\ref{fig:metric_times} concerns metric-computation cost (offline analysis), whereas Fig.~\ref{fig:JRQTvsJQT} and Table~\ref{tab:res_cons_energy} report decoder runtime during playback.}

\subsection{Decoding Energy Consumption Analysis}
\label{sec:dec_energy_measure}
\begin{figure}[t]
    \centering
    \begin{subfigure}{0.61\linewidth}
        \includegraphics[width=\linewidth]{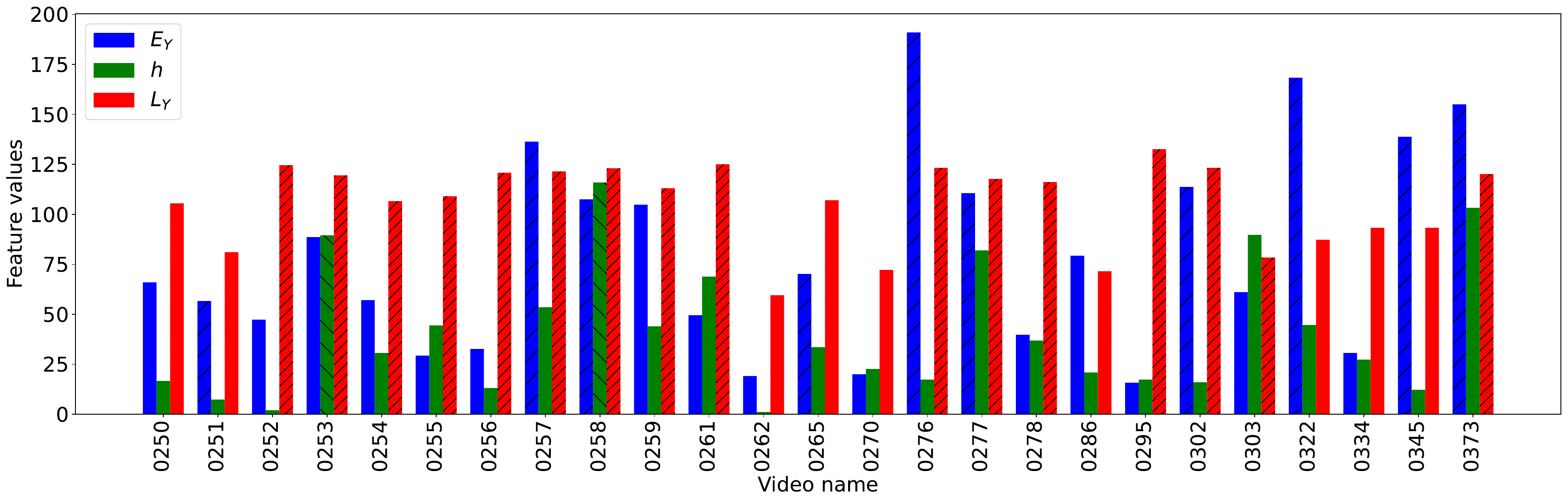}
    \caption{Content complexity features of the representative sequences.}
     %\vspace{.5em}
     \label{fig:vca_res}
    \end{subfigure}
    \begin{subfigure}{0.35\linewidth}
       \includegraphics[width=\linewidth]{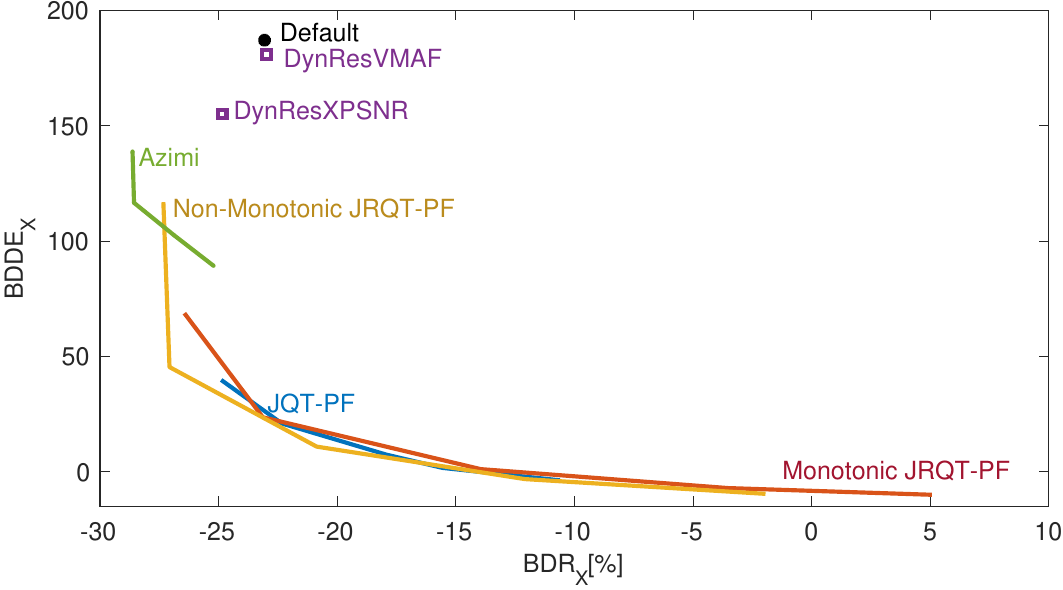} 
       \caption{$BDR_X$-$BDDE_X$ tradeoffs.}
       %\vspace{-.5em}
       \label{fig:bddex_bdrx}
    \end{subfigure}
    \caption{Comparison of energy consumption against compression gains over \fixedladder \vig{for the representative sequences used for the decoding energy consumption analysis}. Lower $BDDE_X$ and $BDR_X$ indicate best tradeoffs. }
    %\vspace{-1.5em}
    \label{fig:sampled_EhL}
\end{figure}

There is a strong linear correlation of decoding time with the energy consumed at decoding, as illustrated in Fig.~\ref{fig:energy_time_corr}. To assess the impact of the proposed multi-objective PF optimization strategies on the energy consumed, we selected 25 sequences from the Inter-4K dataset using K-means clustering based on the VCA content features (\EY, \h, \LY), as shown in Fig.~\ref{fig:vca_res}. The decoding time and energy measurements were performed on an Intel Core i7-12800H laptop. \rev{While relative reductions in decoding complexity are expected to generalize qualitatively across devices, absolute values may vary depending on hardware architecture (\eg ARM vs. x86), operating system, or playback implementation. Investigating cross-device consistency remains a crucial avenue for future research.}

Fig.~\ref{fig:bddex_bdrx} confirms trends similar to $BDR_X \textrm{-} \Delta T_D$ trade-offs. The multi-objective strategies can offer parameterizable tradeoffs in the energy-quality-rate space. Notably $BDDE_X$ gains (\mbox{$BDDE_X<0$}) can be achieved over the \fixedladder whilst ensuring BD rate gains (\mbox{$BDR_X<0$}) for $\alpha_M\geq 0.75$ for \jrqtpf strategies and $\alpha_M\geq 2.5$ for \jqtpf. Azimi~\etal method offers high BD rate gains at the cost of higher energy consumption.

\subsection{Limitations and Future Work}
\vig{While our evaluation uses XPSNR and VMAF as perceptual metrics, future work should include formal subjective studies such as JND testing or MOS evaluation to validate perceptual gains across the proposed ladders. This would strengthen the connection between objective optimization and actual user experience. Furthermore, we focused on offline optimization of bitrate ladders for adaptive video streaming, where the representation set is computed in advance based on content characteristics and decoder-side constraints. A limitation of this approach is that it does not account for the inherent unpredictability of network bandwidth during playback. In real-world streaming, particularly for live or mobile scenarios, optimal ladder selection may not translate to optimal runtime behavior due to fluctuating downlink capacity. Prior research has proposed reinforcement learning (RL)-based~\cite{pensive_ref} and model predictive control (MPC)-based~\cite{mpc_ref1} methods that adapt to network conditions at the client side. More recent systems, such as COMyCo~\cite{comyco_ref}, also consider memory and compute constraints, enabling smarter adaptation across a wider range of devices and network conditions. While these approaches dynamically select the best representation from a predefined ladder, our method is designed to generate a \emph{better ladder to begin with}-- that captures trade-offs between quality, bitrate, and decoding complexity. These two approaches are thus complementary.}

\rev{Another limitation concerns the encoding side. Like all adaptive encoding frameworks, our solution requires encoding each sequence across multiple resolutions and QPs to populate the RQT space, which imposes a higher computational cost compared to fixed ladder baselines that only encode one QP per resolution. However, this overhead is incurred only once during offline ladder construction and is consistent with prior convex-hull approaches~\cite{netflix_paper,Katsenou_vmaf_ladder}. Future work could reduce this cost by leveraging partial encoding, scalable coding, or ML-based prediction of anchor points~\cite{jtps_ref,menon2024convexhull_xpsnr}.}

\rev{Furthermore, our evaluation was performed on a single hardware platform. However, we have shown proportional trends in decoding time and energy, which are expected to hold across devices (\cf Fig.~\ref{fig:energy_time_corr}). Validating these results on diverse architectures (\eg, mobile ARM processors or hardware decoders) is left as future work. Furthermore, the effects of external factors such as CPU frequency scaling, thermal throttling, and background processes, which may affect energy usage beyond what decoding time alone captures, need to be investigated. We would also like to note that although decoding time was used as a proxy for device-side feasibility, end-user QoE also depends on multiple factors such as latency, rebuffering, and playback smoothness. These were outside the scope of this work. Future extensions could integrate our monotonic, complexity-aware ladders into client-side adaptation frameworks that explicitly account for such QoE dimensions.}

\vig{Finally, as future work, we plan to integrate our Pareto-optimized ladders into online decision frameworks, enabling client-side agents to make bandwidth-aware selections from a perceptually and computationally efficient ladder.}

\section{Conclusions}
\label{sec: Conclusions}
This paper introduced a multi-objective Pareto-front optimization framework tailored for adaptive streaming of VVC-encoded content. It addresses the critical trade-offs among bitrate, video quality, and decoding complexity. Our approach enables the construction of bitrate ladders that are perceptually consistent, resource-aware, and adaptable to diverse deployment scenarios. In particular, incorporating quality monotonicity ensures smooth adaptation across resolutions and prevents disruptive quality fluctuations that may negatively impact user experience. We demonstrated that the proposed PF-based strategies flexibly regulate the bitrate-quality-decoding complexity spectrum, adapting to different streaming goals, such as maximizing quality under complexity constraints or minimizing decoding time while preserving perceptual fidelity. The experimental evaluation on the Inter-4K dataset showed that our method substantially reduces bitrate and decoding time compared to fixed and state-of-the-art adaptive ladder construction methods, without compromising perceived video quality. Furthermore, using decoding time as a proxy for complexity enables sustainable high-quality streaming, especially on energy-constrained client devices. Although our framework is currently optimized for offline VOD scenarios, it is compatible with future online adaptive schemes. Future directions include integrating PF-based ladders into reinforcement learning-based adaptation agents and extending the methodology to mixed-codec and cross-device streaming environments.

Overall, the proposed framework provides a principled, extensible foundation for codec-aware, complexity-constrained adaptive streaming, particularly in the context of next-generation video compression standards like VVC.

%\section{Conclusions}
%\label{sec: Conclusions}
%This paper introduced a multi-objective Pareto-front optimization approach for adaptive VVC streaming, addressing the critical trade-offs among bitrate, video quality, and decoding complexity. %We introduced a multi-objective Pareto-front optimization approach for adaptive VVC streaming, addressing the vital tradeoffs among bitrate, video quality, and decoding complexity.
%Notably, the quality-monotonic bitrate ladders constructed by our framework mitigate disruptions caused by quality degradations, ensuring a high QoE. The experimental results showed that the proposed method offers a flexible regulation of the bitrate-quality-decoding complexity tradeoff, making it suitable for various on-demand video streaming use cases with different requirements. Furthermore, by integrating decoding time as a proxy for energy consumption, our proposed framework enables sustainable, high-quality video streaming tailored to the demands of energy-constrained and battery-powered devices.

%Future work aims to perform a subjective study to verify the perceived QoE of the proposed ladders. \vig{We will explore incorporating the proposed PF-based ladders into online adaptive streaming agents using reinforcement learning or predictive control, allowing dynamic selection based on network feedback while preserving perceptual and energy-efficiency constraints.} Finally, we will evaluate playback stability, buffering impact, or DASH/HLS adaptation behavior.

\bibliographystyle{ACM-Reference-Format}
\bibliography{references}
%\bibliography{references}{}
%\bibliographystyle{IEEEtran}
%\input{sec_biography}
\balance
\balance
\end{document}